\documentclass[twocolumn,conference]{IEEEtran}

\usepackage{graphicx,epic,eepic,epsfig,amsmath,latexsym,amssymb,verbatim,subfigure,color}
\usepackage{theorem}
\usepackage{url}
\usepackage{color}

\newcommand{\ket}[1]{|{#1}\rangle}
\newcommand{\bra}[1]{\langle{#1}|}

\newcommand{\iden}{1 \hspace{-1.0mm}  {\bf l}}

\newcommand{\ncd}{\newcommand}
\ncd{\QC}{$\mbox{QC}_{\cal{C}}\;$}
\ncd{\QCpr}{${\mbox{QC}_{\cal{C}}}^\prime\;$}
\ncd{\QCns}{$\mbox{QC}_{\cal{C}}$}
\ncd{\QCprns}{${\mbox{QC}_{\cal{C}}}^\prime$}
\ncd{\cskN}{{|\phi_{\{\kappa\} } \rangle}_{{\cal{C}}_N}}
\ncd{\cskNpr}{{|\phi_{\{\kappa^\prime\} } \rangle}_{{\cal{C}}_N}}
\ncd{\cskNtil}{{|\phi_{\{\tilde{\kappa} \} } \rangle}_{{\cal{C}}_N}}
\ncd{\csk}{{|\phi_{\{\kappa\} } \rangle}_{\cal{C}}}
\ncd{\csktil}{{|\phi_{\{\tilde{\kappa} \} } \rangle}_{\cal{C}}}
\ncd{\cskf}{|\phi_{\{\kappa\} } \rangle_{\cal{C}}}
\ncd{\csktilf}{|\phi_{\{\tilde{\kappa} \} } \rangle_{\cal{C}}}
\ncd{\bracsk}{\mbox{}_{\cal{C}}\langle\phi_{\{\kappa\} }|}
\ncd{\bracsktil}{\mbox{}_{\cal{C}}\langle\phi_{\{\tilde{\kappa} \} }|}
\ncd{\nbracsk}{\mbox{}_{\cal{C}}\langle\phi_{\{\kappa\} }}
\ncd{\nbracsktil}{\mbox{}_{\cal{C}}\langle\phi_{\{\tilde{\kappa} \} }}
\ncd{\cs}{|\phi \rangle_{\cal{C}}\;}
\ncd{\csns}{|\phi \rangle_{\cal{C}}}
\ncd{\nbgh}{\text{nbgh}}
\ncd{\Sab}{S^{ab}}
\ncd{\Sba}{S^{ba}}
\ncd{\ds}{\displaystyle}
\ncd{\ovl}{\overline}

\newtheorem{example}{Example}
\newtheorem{lemma}{Lemma}
\newtheorem{theorem}{Theorem}

\newtheorem{remark}{Remark}
\newtheorem{proposition}{Proposition}

\newcommand{\nc}{\newcommand}
\nc{\rnc}{\renewcommand}
\nc{\beq}{\begin{equation}}
\nc{\eeq}{{\end{equation}}}
\nc{\beqa}{\begin{eqnarray}}
\nc{\eeqa}{\end{eqnarray}}
\nc{\lbar}[1]{\overline{#1}}
\nc{\ketbra}[2]{|#1\rangle\!\langle#2|}
\nc{\braket}[2]{\langle#1|#2\rangle}
\nc{\proj}[1]{| #1\rangle\!\langle #1 |}
\nc{\avg}[1]{\langle#1\rangle}
\nc{\Rank}{\operatorname{Rank}}
\nc{\smfrac}[2]{\mbox{$\frac{#1}{#2}$}}
\nc{\Tr}{\operatorname{Tr}}
\nc{\id}{\operatorname{id}}
\nc{\1}{\openone}
\nc{\ox}{\otimes}
\nc{\dg}{\dagger}
\nc{\dn}{\downarrow}
\nc{\cA}{{\cal A}}
\nc{\cB}{{\cal B}}
\nc{\cC}{{\cal C}}
\nc{\cD}{{\cal D}}
\nc{\cE}{{\cal E}}
\nc{\cF}{{\cal F}}
\nc{\cG}{{\cal G}}
\nc{\cH}{{\cal H}}
\nc{\cI}{{\cal I}}
\nc{\cJ}{{\cal J}}
\nc{\cK}{{\cal K}}
\nc{\cL}{{\cal L}}
\nc{\cM}{{\cal M}}
\nc{\cN}{{\cal N}}
\nc{\cO}{{\cal O}}
\nc{\cP}{{\cal P}}
\nc{\cR}{{\cal R}}
\nc{\cS}{{\cal S}}
\nc{\cT}{{\cal T}}
\nc{\cX}{{\cal X}}
\nc{\cY}{{\cal Y}}
\nc{\cZ}{{\cal Z}}
\nc{\var}{\operatorname{var}}
\nc{\rar}{\rightarrow}
\nc{\lrar}{\longrightarrow}
\nc{\polylog}{\operatorname{polylog}}

\nc{\RR}{{{\mathbb R}}}
\nc{\CC}{{{\mathbb C}}}
\nc{\FF}{{{\mathbb F}}}
\nc{\NN}{{{\mathbb N}}}
\nc{\ZZ}{{{\mathbb Z}}}
\nc{\PP}{{{\mathbb P}}}
\nc{\QQ}{{{\mathbb Q}}}
\nc{\UU}{{{\mathbb U}}}
\nc{\EE}{{{\mathbb E}}}
\nc{\Icoh}{{I^{\rm coh}}}
\nc{\Qca}{{Q_{\rm ss}}}
\nc{\Qcaa}{{Q^{(1)}_{\rm ss}}}
\nc{\Dcaa}{{D^{(1)}_{{\rm ss}\rightarrow}}}
\nc{\Dca}{{D_{{\rm ss}\rightarrow}}}

\nc{\be}{\begin{equation}}
\nc{\ee}{{\end{equation}}}
\nc{\bea}{\begin{eqnarray}}
\nc{\eea}{\end{eqnarray}}
\nc{\<}{\langle}
\rnc{\>}{\rangle}
\nc{\Hom}[2]{\mbox{Hom}(\CC^{#1},\CC^{#2})}
\nc{\rU}{\mbox{U}}

\begin{document}

\title{Multi-Error-Correcting Amplitude Damping Codes}

\author{\IEEEauthorblockN{Runyao Duan\IEEEauthorrefmark{1}\IEEEauthorrefmark{2}, 
Markus Grassl\IEEEauthorrefmark{3}, 
Zhengfeng Ji\IEEEauthorrefmark{4}, and
Bei Zeng\IEEEauthorrefmark{5}}
\IEEEauthorblockA{\IEEEauthorrefmark{1}Centre for Quantum Computation and Intelligent Systems (QCIS),\\
Faculty of Engineering and Information Technology, University of Technology, Sydney, NSW 2007, Australia}
\IEEEauthorblockA{\IEEEauthorrefmark{2}State Key Laboratory of
  Intelligent Technology and Systems, Tsinghua National Laboratory for
  Information Science\\ and Technology, Department of Computer Science
  and Technology, Tsinghua University, Beijing 100084, China}
\IEEEauthorblockA{\IEEEauthorrefmark{3}Centre for Quantum
  Technologies, National University of Singapore, Singapore 117543, Singapore}
\IEEEauthorblockA{\IEEEauthorrefmark{4}Perimeter Institute for Theoretical Physics, Waterloo, ON, N2L2Y5, Canada}
\IEEEauthorblockA{\IEEEauthorrefmark{5}Institute for Quantum Computing
  and the Department of Combinatorics and Optimization,\\
  University of Waterloo, Waterloo, ON, N2L3G1, Canada}}

\maketitle
\begin{abstract}
We construct new families of multi-error-correcting quantum codes for the
amplitude damping channel.  Our key observation is that, with proper
encoding, two uses of the amplitude damping channel simulate a quantum
erasure channel. This allows us to use concatenated codes with quantum
erasure-correcting codes as outer codes for correcting multiple
amplitude damping errors.  Our new codes are degenerate stabilizer codes 
and have parameters which are better than the amplitude damping 
codes obtained by any previously known construction.
\end{abstract}


%%%%%%%%%%%%%%%%%%%%%%%%%%%%%%%%%%%%%%%%%%%%%%%%%%%%%%%%%%%%%%%%%%%%%%%%%%%%%

\section{Introduction}

In most of works on quantum error correction, it is assumed that
the errors to be corrected are completely random, with no knowledge
other than that they affect different qubits independently
\cite{nielsenchuang,thesis:gottesman}.  Or, equivalently, this is to
assume that the Pauli-type errors
$X=\left(\begin{smallmatrix}0&1\\1&0\end{smallmatrix}\right)$, 
$Y=\left(\begin{smallmatrix}0&-i\\i&0\end{smallmatrix}\right)$, and
$Z=\left(\begin{smallmatrix}1&0\\0&-1\end{smallmatrix}\right)$, 
happen with equal probability $p_x=p_y=p_z=p/3$.  The quantum channel
described by this kind of noise is called depolarizing channel
$\mathcal{E}_{DP}$.

The most general physical operations (or quantum channels) allowed by
quantum mechanics are completely positive, trace preserving linear
maps which can be represented in the following Kraus decomposition
form:
\begin{equation}
\cN(\rho)=\sum_k A_k \rho A_k^\dag,
\end{equation}
where $A_k$ are called Kraus operators of the quantum channel $\cN$
and satisfy the completeness condition $\sum_k A_k^\dag A_k=\iden$.
In this language of quantum channels, the depolarizing channel
$\mathcal{E}_{DP}$ with error parameter $p$ acting on any one-qubit quantum
state $\rho\in\CC^{2\times 2}$ as
\begin{equation}
\mathcal{E}_{DP}(\rho)=(1-p)\rho+\frac{p}{3}(X\rho X+Y\rho Y +Z\rho Z),
\end{equation}
so the Kraus operators for the depolarizing channel are the Pauli
matrices together with identity.

However, if further information about an error process is available,
more efficient codes can be designed.  Indeed in many physical
systems, the types of noise are likely to be unbalanced between
amplitude ($X$-type) errors and phase ($Z$-type) errors. Recently a
lot of attention has been put into designing codes for this situation
and in studying their fault tolerance properties
\cite{aliferis-biased-2007,evans-2007,fletcher-ad,Ioffe-bias,Martin:ISIT2008}. All
those works deal with error models which are still described by Kraus
operators that are Pauli matrices (Pauli Kraus operators), but the
$X$- and $Y$-errors happen with equal probability $p_x=p_y$, which
might be different from the probability $p_z$ that a $Z$-error
happens.  The quantum channels described by this kind of noise are
called asymmetric channels $\mathcal{E}_{AS}$ acting on any one-qubit
quantum state $\rho$ as
\begin{alignat}{4}
\mathcal{E}_{AS}(\rho)&{}=(1-(2p_x+p_z))\rho\nonumber\\
&\quad{}+p_x\left(X\rho X+Y\rho Y\right) +p_zZ\rho Z.
\end{alignat}
The choice $p_x=p_y$ is related to a physically realistic error model
including amplitude damping (AD) noise and phase damping noise
\cite{nielsenchuang}.  The Kraus operators for AD noise with damping
rate $\gamma$ are
\begin{equation}
A_0=\begin{pmatrix} 1 & 0 \\0 &
\sqrt{1-\gamma} \end{pmatrix}
\quad\text{and}\quad A_1=\begin{pmatrix} 0 &
\sqrt{\gamma} \\0 & 0 \end{pmatrix}.
\label{eq:ADKraus}
\end{equation}
Note that
\begin{alignat*}{5}
A_1=\begin{pmatrix} 0 &
\sqrt{\gamma} \\0 & 0 \end{pmatrix}&{}=\frac{\sqrt{\gamma}}{2}\left(X+iY\right)\quad\text{and}\\
A_1^\dag&{}=\frac{\sqrt{\gamma}}{2}\left(X-iY\right).
\end{alignat*}
Hence the linear span of the operators $A_1$ and $A_1^\dag$ equals the
linear span of $X$ and $Y$.  If the system is at finite temperature,
the Kraus operator $A^{\dagger}_1$ will appear in the noise model
\cite{nielsenchuang}.  Thus, if the code is capable of correcting $t$
$X$- and $t$ $Y$-errors, it can also correct $t$ $A_1$- and $t$
$A_1^{\dagger}$-errors.

It was observed that when the temperature of a physical system is zero
or very low, the error $A_1^\dag$ is actually negligible
\cite{nielsenchuang}. For simplicity, we further ignore the phase
damping error (which is characterized by the Pauli operator $Z$).
Then the error model is fully characterized by $A_0$ and $A_1$. In
this work, we will focus on this quantum channel with only amplitude
damping noise, i.e. the AD channel $\mathcal{E}_{AD}$, with only two
Kraus operators given by Eq. (\ref{eq:ADKraus}).  The AD channel is
the simplest nonunital channel whose Kraus operators cannot be
described by Pauli operations. The AD channel is a quantum analogue of
the classical $\mathcal{Z}$-channel 
which transmits $0$ faithfully, but maps $1$ to either $0$ or $1$
\cite{GF3quantum}. For the AD channel we only need to deal with the
error $A_1$ (a quantum analogue of the error $1\rightarrow 0$), but
not with $A^{\dagger}_1$ (a quantum analogue of the error
$0\rightarrow 1$).  So asking to be able to correct both $X$- and
$Y$-errors is a less efficient way for constructing quantum codes for
the AD channel.

Since the error model is not described by Pauli Kraus operators, the
task of constructing good error-correcting codes becomes very
challenging.  The known techniques dealing with Pauli errors cannot be
applied or result in codes with bad parameters.  Several new
techniques for the construction of codes which are adapted to this
type of noise with non-Pauli Kraus operators, and the AD channel in
particular, have been developed
\cite{Chuang1997,fletcher-ad,lang-ad,leung-1997,GF3quantum}.  After
years' effort, systematic methods for constructing high performance
single-error-correcting codes have been found
\cite{lang-ad,GF3quantum}.  However, all these methods fail to
construct good AD codes correcting multi-errors.

In this paper we present a method for finding families of codes
correcting multi-amplitude-damping errors.  Our construction is based
on the observation that with respect to a simple encoding two uses of
the amplitude damping channel simulate a quantum erasure channel.
This allows us to apply a concatenated coding scheme with quantum
erasure-correcting codes as outer codes, resulting in codes correcting
multi-amplitude-damping errors.  Our new codes are degenerate
stabilizer codes which have better parameters than the codes given by
any previously known construction.

\section{Correcting amplitude damping errors\label{AD}} 
\label{sec:adcode}

A quantum error-correcting code $Q$ is a subspace of $(\CC^2)^{\otimes
  n}$, the space of $n$ qubits.  For a $K$-dimensional code space
spanned by the orthonormal set $\ket{\psi_i}$, $i=1,\ldots,K$ and a
set of errors $\cE$ there is a physical operation correcting all
elements $E_\mu \in \cE$ if the error correction conditions
\cite{bennett-1996-54,KL97} are satisfied:
\begin{equation}
\forall_{ij,\mu\nu}\quad\bra{\psi_i} E_\mu^\dag E_\nu \ket{\psi_j} = C_{\mu\nu}\delta_{ij}, 
\label{eq:qecccondition}
\end{equation}
where $C_{\mu \nu}$ depends only on $\mu$ and $\nu$. If the matrix
$(C_{\mu \nu})$ has full rank the code is said to be nondegenerate,
otherwise it is degenerate.

For the AD channel, if $\gamma$ is small, we would like to correct the
leading order errors that occur during amplitude damping.
Setting $A=X+iY$ and $B=I-Z$, we have
\begin{equation}
A_1=\frac{\sqrt{\gamma}}{2}A\quad\text{and}\quad A_0=I-\frac{\gamma}{4}B+O(\gamma^2).
\end{equation}
It has been shown that in order to improve the fidelity of the
transmission through an amplitude damping channel from $1-\gamma$ to
$1-\gamma^t$, it is sufficient to satisfy the error-detection
conditions for $2t$ $A$-errors and $t$ $B$-errors \cite[Section
  8.7]{thesis:gottesman}.  We will say that such a code corrects $t$
amplitude damping errors since it improves the fidelity, to leading
order, just as much as a true $t$-error-correcting code would for the
same channel.

Stabilizer codes are a large kind of quantum codes which contain many
good quantum codes \cite{thesis:gottesman,nielsenchuang}.  A
stabilizer code with $n$ qubits encoding $k$ qubits is of distance $d$
if all errors of weight at most $d-1$ (i.e., operators acting
nontrivially on less than $d$ individual qubits) can be detected or
have no effect on $Q$, and we denote the parameters of $Q$ by
$[[n,k,d]]$.  We say an $[[n,k]]$ stabilizer code is a $t$-code if it
corrects $t$ AD-errors.  For comparison with stabilizer codes, we say
an $[[n,k]]$ $t$-code is good if $2t+1>d$ for the best possible
$[[n,k,d]]$ code; or, $n<n'$ for the best possible $[[n',k,2t+1]]$
code; or, $k>k'$ for the best possible $[[n,k',2t+1]]$ code.

The first AD code given by Leung et al. \cite{leung-1997} is a
$[[4,1]]$ $1$-code, i.e., correcting a single AD-error.  Basis vectors
of the code are
\begin{alignat}{5}
\ket{0}_{L}&{}=\frac{1}{\sqrt{2}}\left(\ket{0000}+\ket{1111}\right)\nonumber\\
\ket{1}_{L}&{}=\frac{1}{\sqrt{2}}\left(\ket{0011}+\ket{1100}\right).
\label{eq:Leung}
\end{alignat}
Using only 4 qubits, this $1$-code is better than the $[[5,1,3]]$
code, a quantum code correcting an arbitrary single-qubit error and
encoding one qubit using the minimal number of qubits
\cite{bennett-1996-54,PhysRevLett.77.198}.

Following the work by Leung et al. \cite{leung-1997}, several
constructions for $1$-codes have been proposed
\cite{fletcher-ad,thesis:gottesman,lang-ad,GF3quantum}, including some
high performance $1$-codes.  However, very little is known about good
multi-error-correcting AD codes.  It turns out that none of the
methods known for constructing good $1$-codes can be directly
generalized to $t$-codes with $t>1$.

Gottesman \cite[Section 8.7]{thesis:gottesman} has shown that Shor's nine-qubit code \cite{shor95}
\begin{alignat}{5}
\ket{0}_{L}&{}=\frac{1}{2\sqrt{2}}\left(\ket{000}+\ket{111}\right)^{\otimes 3}\nonumber\\
\ket{1}_{L}&{}=\frac{1}{2\sqrt{2}}\left(\ket{000}-\ket{111}\right)^{\otimes 3}
\label{eq:shor}
\end{alignat}
can correct two AD-errors, despite the fact that it can correct only
a single general error.  It is the best known $2$-code and it is better
than the $[[11,1,5]]$ code \cite{thesis:gottesman}, the best
two-error-correcting stabilizer code encoding one qubit
\cite{Grassl:tables}.

It is interesting to note that the $1$-code given by Eq. (\ref{eq:Leung}) can be rewritten
in another basis as
\begin{alignat}{5}
\ket{+}_{L}&{}=\frac{1}{\sqrt{2}}\left(\ket{0}_L+\ket{1}_L\right)=\frac{1}{2}\left(\ket{00}+\ket{11}\right)^{\otimes 2}\nonumber\\
\ket{-}_{L}&{}=\frac{1}{\sqrt{2}}\left(\ket{0}_L-\ket{1}_L\right)=\frac{1}{2}\left(\ket{00}-\ket{11}\right)^{\otimes 2},
\label{eq:Leung2}
\end{alignat}
which is of a similar form as Eq. (\ref{eq:shor}).

Therefore, we can generalize the constructions of
Eqs. (\ref{eq:Leung2}) and (\ref{eq:shor}) to $t$-codes with basis
\begin{alignat}{5}
\ket{0}_{L}&{}={2^{-\frac{t+1}{2}}}\Bigl(\underbrace{\ket{0\ldots 0}}_{t+1}+\underbrace{\ket{1\ldots 1}}_{t+1}\Bigr)^{\otimes(t+1)}\nonumber\\
\ket{1}_{L}&{}={2^{-\frac{t+1}{2}}}\Bigl(\underbrace{\ket{0\ldots 0}}_{t+1}-\underbrace{\ket{1\ldots 1}}_{t+1}\Bigr)^{\otimes(t+1)}.
\label{eq:shorge}
\end{alignat}
However, these $[[n^2,1,n]]$ so-called Bacon-Shor code
\cite{bacon-sub-2005,shor95} correcting $t=n-1$ AD-errors scale badly
when $n$ is large.  For instance, there exists a $[[25,1,9]]$ code and
a $[[29,1,11]]$ code \cite{Grassl:tables}.

Note that these $[[n^2,1]]$ codes are of Calderbank-Shor-Steane (CSS)
type \cite{calderbank96,steane96}.  They are also degenerate: for
instance, a $Z$-error acting on the first qubit or the second qubit
has the same effect on the code.

In general, CSS codes can be used to construct codes for the AD
channel \cite[Section 8.7]{thesis:gottesman}: 
\begin{proposition}
An $[[n,k]]$ CSS code of $X$-distance $2t+1$ and $Z$-distance $t+1$ is
an $[[n,k]]$ $t$-code.
\end{proposition}

In the first column of Table \ref{tab:CSSbounds} we provide bounds on
the length $n$ of codes for the AD channel encoding one or two qubits
derived from CSS codes with given $Z$- and $X$-distances $t+1$ and
$2t+1$, respectively.  The lower bounds have been derived using linear
programming techniques \cite{Sarvepalli}.  The upper bound is based on
CSS codes constructed from the database of best known linear codes
\cite{Magma,Grassl:tables}.

In the fifth column we give upper and lower bounds on the length $n'$
such that an $[[n',k,t+1]]$ code may exist.  In the last column, we
list the bounds on the length of $t$-code from Theorem \ref{th}.  The
data for columns $n'$ and $2m$ is taken from \cite{Grassl:tables}).

\begin{table}[hbt]
\begin{center}
\begin{tabular}{c c c c c c}
\hline
$n$ & $k$ & $t+1$ & $2t+1$ & $n'$ & $2m$\\
\hline
12--13 & 1 &  3 &  5 & 11     & 10     \\
19--20 & 1 &  4 &  7 & 17     & 20     \\
25--30 & 1 &  5 &  9 & 23--25 & 22     \\
33--41 & 1 &  6 & 11 & 29     & 32     \\
39--54 & 1 &  7 & 13 & 35--43 & 34     \\
47--70 & 1 &  8 & 15 & 41--53 & 44--48 \\
53--79 & 1 &  9 & 17 & 47--61 & 46--50 \\
  --89 & 1 & 10 & 19 & 53--81 & 56     \\
  --105& 1 & 11 & 21 & 59--85 & 58     \\
\hline
14--17 & 2 &  3 &  5 & 14      & 16     \\ 
20--27 & 2 &  4 &  7 & 20--23  & 20     \\ 
27--37 & 2 &  5 &  9 & 26--27  & 28     \\ 
34--45 & 2 &  6 & 11 & 32--41  & 32     \\ 
41--62 & 2 &  7 & 13 & 38--51  & 40--46 \\ 
  --71 & 2 &  8 & 15 & 44--59  & 44--52 \\
  --87 & 2 &  9 & 17 & 50--78  & 52--54 \\
  --102& 2 & 10 & 19 & 56--83  & 56--56 \\
  --110& 2 & 11 & 21 & 62--104 & 64--82 \\
\hline
\end{tabular}
\end{center}
\caption{Bounds on the length $n$ of an $[[n,1]]$ $t$-code derived
  from CSS codes, together with the bounds on the length $n'$ of a
  stabilizer code $[[n',1,2t+1]]$ and the length $2m$ of an $[[2m,1]]$
  $t$-code from Theorem \ref{th}.\label{tab:CSSbounds}}
\end{table}

It can be seen from Table \ref{tab:CSSbounds} that the construction of
AD codes based on CSS codes unlikely gives good AD codes.  But as it is unknown
whether these bounds for $n$ and $n'$ given in this table can be
achieved, we do not have the definite answer. This problem will be
addressed in future research.

\section{AD code based on quantum erasure codes}
\label{sec:aderasure}
As discussed in Sec. \ref{sec:adcode}, no good method is known for
constructing good multi-error-correcting AD codes.  In this section we
provide a construction which systematically gives high performance
$t$-codes with $t>1$.  The construction uses concatenated quantum
codes with an inner and an outer quantum code.  After decoding the
inner quantum code, the effective channel is a quantum erasure
channel.  We start by proving the following lemma.

\begin{lemma}\label{lemma:q_dual_rail}
Using the quantum \emph{dual-rail code} $\mathcal{Q}_i$ which encodes a
single qubit into two qubits, given by
\begin{equation}
\ket{0}_L=\ket{01},\quad \ket{1}_L=\ket{10},
\label{eq:inner}
\end{equation}
two uses of the AD channel simulate a quantum erasure channel.
\end{lemma}
\begin{IEEEproof}
For any state $\rho$ of the code $\mathcal{Q}_1$, we observe that
\begin{equation}
\mathcal{E}_{AD}^{\otimes 2}(\rho)=(1-\gamma)\rho+\gamma(\ket{00}\bra{00}).
\end{equation}
The state $\ket{00}$ is orthogonal to the code $\mathcal{Q}_1$. Using
a measurement that either projects on $\mathcal{Q}_1$ or its
orthogonal complement, it can be detected whether an AD error occurred
or not.  Hence we obtain a quantum erasure channel with erasure symbol
$\ket{00}$.
\end{IEEEproof}

\begin{remark}
It can easily be shown that with respect to the dual-rail code
$\{01,10\}$, two uses of the $\mathcal{Z}$-channel simulate a
classical erasure channel with erasure symbol $00$ (see,
e.g. \cite{Massey}). Lemma \ref{lemma:q_dual_rail} is a quantum
analogue of this fact, yet Lemma \ref{lemma:q_dual_rail} is nontrivial
due to the Kraus operator $A_0$, which introduces some relative phase
error between $\ket{0}$ and $\ket{1}$ that has no classical analogue.
\end{remark}

Lemma \ref{lemma:q_dual_rail} allows us to use quantum
erasure-correcting codes as outer codes for correcting multiple
amplitude damping errors.  It is known that an $[[m,k,d]]$ quantum
code corrects $d-1$ erasure errors
\cite{thesis:gottesman,grassl97,nielsenchuang}. Our main result is
given by the following theorem.
\begin{theorem}
If there exists an $[[m,k,d]]$ quantum code, then
there is a $[[2m,k]]$ code correcting $t=d-1$ amplitude damping errors.
\label{th}
\end{theorem}
\begin{IEEEproof}
Let $\mathcal{Q}$ be the concatenated code with the inner code
$\mathcal{Q}_1$ given Eq. (\ref{eq:inner}) and the outer
code $\mathcal{Q}_2$ with parameters $[[m,k,d]]$.  The code
$\mathcal{Q}_2$ corrects $d-1$ erasure errors.  A single AD-error on
each block of the inner code creates an erasure error for the outer
code.  The position of the error is indicated by the erasure state
$\ket{00}$.  Hence the outer codes takes care of $d-1$ AD-errors
acting on different blocks.  Two errors acting on the same block
annihilate the state, such that the quantum error correction condition
given by Eq. (\ref{eq:qecccondition}) is naturally satisfied. Hence
$\mathcal{Q}$ is a $[[2m,k]]$ AD code correcting $t=d-1$ amplitude
damping errors.
\end{IEEEproof}

\begin{remark}
It is interesting to compare our construction with the corresponding
classical case, where concatenation with the dual-rail code
$\{01,10\}$ as inner code and an $[m,k,d]$ erasure-correcting code as
outer code yields an $[2m,k]$ $(d-1)$-code for the
$\mathcal{Z}$-channel.  However, this $(d-1)$-code is in general not
good because simply repeating each codeword of an $[m,k,d]$ classical
code will straightforwardly give a $[2m,k,2d]$ code correcting $d-1$
arbitrary errors.  In the quantum case, however, the existence of an
$[[m,k,d]]$ stabilizer code does not necessarily lead to a
$[[2m,k,2d]]$ stabilizer code.
\end{remark}

In Table \ref{tab:t-codes}, we compare the $t$-codes from our
construction with the known upper and lower bounds on the minimum
distance of stabilizer codes from \cite{Grassl:tables}.  We fix the
number of logical qubits $k$ and the number $t$ of correctable
AD-errors within the range $k=1,\ldots,6$ and $t=1,\ldots,10$.  The
length $n=2m$ of the code is derived from the shortest known
stabilizer code with parameters $[[m,k,t+1]]$ from
\cite{Grassl:tables}. Hence the first three columns gives the
parameters of each line in the table corresponds to an $[[n,k]]$
$t$-code.  The fourth column provides $2t+1$, which is the distance
that is required for an $[[n,k]]$ code to be capable to correct $t$
arbitrary errors.  The last column gives the lower and upper bounds on
the distance $d$ of a $[[n,k,d]]$ stabilizer code from
\cite{Grassl:tables}.  Hence all $t$-codes with $2t+1>d$ are better
than the stabilizer codes with the same length and dimension.  With
the exception of small parameters, many of our codes outperform the
known---or even the best possible---corresponding stabilizer codes
correcting $t$ arbitrary errors.   Note that any improvement of the
lower bound on the distance $d$ of a stabilizer code implies some
improvement for $t$-codes as well.  

\begin{table*}[bht]
\begin{center}
\begin{tabular}{c@{\kern8mm}c@{\kern8mm}c}
\begin{tabular}{c c c c c}
\hline
$n$ & $k$ & $t$ & $2t+1$ & $d$\\
\hline
 8 &  1 &  1 &  3 &        3 \\
10 &  1 &  2 &  5 &        4 \\
20 &  1 &  3 &  7 &        7 \\
22 &  1 &  4 &  9 &     7--8 \\
32 &  1 &  5 & 11 &       11 \\
34 &  1 &  6 & 13 &   11--12 \\
48 &  1 &  7 & 15 &   13--17 \\
50 &  1 &  8 & 17 &   13--17 \\
56 &  1 &  9 & 19 &   15--19 \\
58 &  1 & 10 & 21 &   15--20 \\
\hline
 8 &  2 &  1 &  3 &        3 \\
16 &  2 &  2 &  5 &        6 \\
20 &  2 &  3 &  7 &     6--7 \\
28 &  2 &  4 &  9 &       10 \\
32 &  2 &  5 & 11 &   10--11 \\
46 &  2 &  6 & 13 &   12--16 \\
52 &  2 &  7 & 15 &   14--18 \\
54 &  2 &  8 & 17 &   14--18 \\
56 &  2 &  9 & 19 &   14--19 \\
82 &  2 & 10 & 21 &   18--28 \\
\hline
\end{tabular}
&
\begin{tabular}{c c c c c}
\hline
$n$ & $k$ & $t$ & $2t+1$ & $d$\\
\hline
12 &  3 &  1 &  3 &        4 \\
16 &  3 &  2 &  5 &        5 \\
24 &  3 &  3 &  7 &     7--8 \\
30 &  3 &  4 &  9 &    9--10 \\
40 &  3 &  5 & 11 &   10--13 \\
48 &  3 &  6 & 13 &   11--16 \\
52 &  3 &  7 & 15 &   13--17 \\
54 &  3 &  8 & 17 &   13--18 \\
72 &  3 &  9 & 19 &   15--24 \\
82 &  3 & 10 & 21 &   18--27 \\
\hline
12 &  4 &  1 &  3 &        4 \\
20 &  4 &  2 &  5 &        6 \\
24 &  4 &  3 &  7 &     6--8 \\
32 &  4 &  4 &  9 &    8--10 \\
40 &  4 &  5 & 11 &   10--13 \\
50 &  4 &  6 & 13 &   12--16 \\
52 &  4 &  7 & 15 &   12--17 \\
70 &  4 &  8 & 17 &   15--23 \\
80 &  4 &  9 & 19 &   16--26 \\
96 &  4 & 10 & 21 &   18--31 \\
\hline
\end{tabular}
&
\begin{tabular}{c c c c c}
\hline
$n$ & $k$ & $t$ & $2t+1$ & $d$\\
\hline
16 &  5 &  1 &  3 &     4--5 \\
22 &  5 &  2 &  5 &     6--7 \\
28 &  5 &  3 &  7 &     7--9 \\
36 &  5 &  4 &  9 &    8--11 \\
42 &  5 &  5 & 11 &    9--13 \\
50 &  5 &  6 & 13 &   11--16 \\
60 &  5 &  7 & 15 &   13--19 \\
78 &  5 &  8 & 17 &   15--25 \\
86 &  5 &  9 & 19 &   18--28 \\
98 &  5 & 10 & 21 &   19--32 \\
\hline
16 &  6 &  1 &  3 &        4 \\
24 &  6 &  2 &  5 &     6--7 \\
28 &  6 &  3 &  7 &     6--8 \\
36 &  6 &  4 &  9 &    8--11 \\
48 &  6 &  5 & 11 &   10--15 \\
58 &  6 &  6 & 13 &   12--19 \\
64 &  6 &  7 & 15 &   14--21 \\
84 &  6 &  8 & 17 &   17--27 \\
92 &  6 &  9 & 19 &   18--29 \\
104 &  6 & 10 & 21 &   19--33 \\
\hline
\end{tabular}
\end{tabular}
\end{center}
\caption{Comparison of our $[[n,k]]$ $t$-codes and the bounds on the
  minimum distance $d$ of a stabilizer code $[[n,k,d]]$.\label{tab:t-codes}}
\end{table*}

Note that all the $t$-codes listed in the table are degenerate
stabilizer codes obtained by concatenation of a stabilizer code as
outer code and the quantum dual-rail code $\mathcal{Q}_1$ given by
Eq. (\ref{eq:inner}) as inner code.  In order to compute the
stabilizer of the concatenated code, note that the inner code
$\mathcal{Q}_1$ is stabilized by $-ZZ$, and has logical operators
$\bar{X}=XX$ and $\bar{Z}=ZI$. As an example, we compute the
stabilizer for the $[[10,1]]$ $2$-code.
\begin{example}
A $[[10,1]]$ $2$-code can be derived from the $[[5,1,3]]$ code with
stabilizer generated by:
\begin{equation}\label{eq:stab_5_1_3}
\begin{array}{r c c @{\;} c @{\;} c @{\;} c @{\;} c}
g_1 &=& X & Z & Z & X & I \\
g_2 &=& I & X & Z & Z & X \\
g_3 &=& X & I & X & Z & Z \\
g_4 &=& Z & X & I & X & Z
\end{array}
\end{equation}
The stabilizer of the $[[10,1]]$ $2$-code
is obtained by replacing the operators in Eq. (\ref{eq:stab_5_1_3}) by
the logical operators of $\mathcal{Q}_1$ and adding the stabilizer for
each block of the inner code:
\begin{equation}\label{eq:stab_10_1}
\begin{array}{r c r@{\,}c @{\;\;} c@{\,}c @{\;\;} c@{\,}c @{\;\;} c@{\,}c @{\;\;} c@{\,}c}
g'_1 &=&  X & X & Z & I & Z & I & X & X & I & I \\
g'_2 &=&  I & I & X & X & Z & I & Z & I & X & X \\
g'_3 &=&  X & X & I & I & X & X & Z & I & Z & I \\
g'_4 &=&  Z & I & X & X & I & I & X & X & Z & I \\[0.75ex]
g'_5 &=& -Z & Z & I & I & I & I & I & I & I & I \\
g'_6 &=& -I & I & Z & Z & I & I & I & I & I & I \\
g'_7 &=& -I & I & I & I & Z & Z & I & I & I & I \\
g'_8 &=& -I & I & I & I & I & I & Z & Z & I & I \\
g'_9 &=& -I & I & I & I & I & I & I & I & Z & Z
\end{array}
\end{equation}
As a degenerate stabilizer code, this code has parameters
$[[10,1,4]]$.  As a $2$-code, this code is not as good as Shor's
nine-qubit code given in Eq. (\ref{eq:shor}), but still better than
the shortest stabilizer code $[[11,1,5]]$ encoding one qubit and
correcting two arbitrary errors.
\end{example}
However, the $[[22,1]]$ $4$-code given in Table \ref{tab:t-codes} is
better than the $[[25,1]]$ $4$-code given in Eq. (\ref{eq:shorge}),
the degenerate $[[25,1,9]]$ code constructed from concatenating two
$[[5,1,3]]$ codes, and even the putative stabilizer code $[[22,1,8]]$.

From the last column in Table~\ref{tab:CSSbounds} we see that, with the
exception when both parameters $t$ and $k$ are small, the codes from
our construction are better than the $t$-codes derived from CSS codes.

\section{Possible Generalizations}
One possible generalization of our construction is to chose a
different inner code.  For instance, we can take the inner code as the
following quantum code $\mathcal{Q}'_1$ which encodes one qutrit into
three qubits:
\begin{equation}
\ket{0}_L=\ket{001},\quad \ket{1}_L=\ket{010},\quad \ket{2}_L=\ket{100}.
\end{equation}

For any state $\rho$ of the code $\mathcal{Q}'_1$, we observe that
\begin{equation}
\mathcal{E}_{AD}^{\otimes 3}(\rho)=(1-\gamma)\rho+\gamma(\ket{000}\bra{000}),
\end{equation}
hence the effective channel is a qutrit quantum erasure channel where
the state $\ket{000}$ indicates an erasure.

Since the inner code $\mathcal{Q}'_1$ is of dimension $3$, the outer
code $\mathcal{Q}'_2$ must be chosen from quantum codes constructed
for qutrits rather than qubits, i.e. $\mathcal{Q}'_2$ is a subspace of
$(\CC^3)^{\otimes m}$.  Using a $[[m,k,d]]_3$ quantum code
$\mathcal{Q}'_2$ (where the subscript $3$ indicates that this is a
qutrit code), the concatenated code $\mathcal{Q}$ with inner code
$\mathcal{Q}_1$ and outer code $\mathcal{Q}_2$ is an AD code
correcting $t=d-1$ AD errors, with length $3m$ and encoding a space of
dimension $3^k$.  In general, quantum code of length $n$ and dimension
$K$ is denoted by $((n,K))$, so this construction yields a
$((3m,3^k))$ $(d-1)$-code.

For instance, an $[[8,2,4]]_3$ outer code (see
\cite{yu-graphical-nonbinary,looi-code}) gives a $((24,9))$ AD code
correcting $3$ AD errors.  This is better than the parameters
$[[24,3,7\text{--}8]]$ of a stabilizer code
(cf. \cite{Grassl:tables}), but worse than the $[[24,4]]$ $3$-code
given in Table \ref{tab:t-codes}.  It is not yet clear whether this or
other generalizations based on concatenation using codes for the
erasure channel yield better AD codes than those obtained from the
quantum dual-rail codes.

\section{Conclusions}
We have constructed families of good multi-error-correcting quantum
codes for the amplitude damping channel based on code concatenation and
quantum erasure-correcting codes. As the rate of our codes can never
exceed the rate $1/2$ of the inner code, other methods---possibly
generalized concatenation of quantum codes
\cite{GCQC:PRA,GCQC:ISIT}---have to be used in order to construct
high-rate AD codes.  However, our method provides the first systematic
construction for good multi-error-correcting AD codes. We hope that our method shade lights
on constructing good quantum codes adapted for other non-Pauli
channels beyond the AD channel, and further understanding on the role
that degenerate codes play in quantum coding theory.

\section*{Acknowledgments}
We thank Daniel Gottesman and Peter Shor for helpful discussions.
RD is partly supported by QCIS, University of Technology, Sydney, and
the NSF of China (Grant Nos. 60736011 and 60702080).
BZ is supported by NSERC and QuantumWorks.
Centre for Quantum Technologies is a Research Centre of Excellence
funded by Ministry of Education and National Research Foundation of
Singapore.
Research at Perimeter Institute is supported by the Government of
Canada through Industry Canada and by the Province of Ontario thought
the Ministry of Research \& Innovation.

\IEEEtriggeratref{10}
\IEEEtriggercmd{\enlargethispage{-10mm}}

\end{document}